\documentclass{article}
\providecommand{\keywords}[1]{\textbf{\textit{Keywords---}} #1}
\usepackage{epsfig}
\usepackage{float}
\usepackage[margin=8.5mm]{caption}
\usepackage[letterpaper, total={7.5in, 9in}]{geometry}

\bibliography{bibliography}

\usepackage{amsmath}
\usepackage{bm}
\usepackage{amsfonts}

\usepackage{setspace}
\newcommand{\Real}{\mathbb{R}}
\newcommand{\dub}{\boldsymbol{w}}
\newcommand{\lb}{\boldsymbol{\ell_b}}
\newcommand{\lp}{\boldsymbol{\ell_p}}
\newcommand{\LETd}{\text{LET}_\text{d}}

\begin{document}

\title{A novel inverse algorithm to solve IPO-IMPT of proton FLASH therapy with sparse filters}
\author{Nathan Harrison, Minglei Kang, Ruirui Liu, Serdar Charyyev, \\
 Niklas Wahl, Wei Liu, Jun Zhou, Kristin A. Higgins, \\
Charles B. Simone II, Jeffrey D. Bradley, William S. Dynan, Liyong Lin}
\maketitle

\noindent {\begin{center} \Large \textbf{Abstract} \end{center}}

\noindent \textbf{Purpose:}
The recently proposed Integrated Physical Optimization Intensity Modulated Proton Therapy (IPO-IMPT) framework allows simultaneous optimization of dose, dose rate, and linear energy transfer (LET) for FLASH treatment planning.
Finding solutions to IPO-IMPT is difficult due to computational intensiveness.
Nevertheless, an inverse solution that simultaneously specifies the geometry of a sparse filter and weights of a proton intensity map is desirable for both clinical and preclinical applications.
Such solutions can reduce effective biological dose to organs at risk in cancer patients as well as reduce the number of animal irradiations needed to derive extra biological dose models in preclinical studies.

\noindent \textbf{Methods:}
Unlike our initial forward heuristic, this inverse IPO-IMPT solution includes simultaneous optimization of sparse range compensation, sparse range modulation, and spot intensity.
The daunting computational tasks vital to this endeavor were resolved iteratively with a distributed computing framework to enable Simultaneous Intensity and Energy Modulation and Compensation (SIEMAC).
SIEMAC was demonstrated on a human central lung cancer patient and a minipig.

\noindent \textbf{Results:}
SIEMAC simultaneously improves maps of spot intensities and patient-field-specific sparse range compensators and range modulators.
For the lung cancer patient, at our maximum nozzle current of 300 nA, dose rate coverage above 100 Gy/s increased from 57\% to 96\% in the lung and from 93\% to 100\% in the heart, and LET coverage above 4 keV/$\mu$m dropped from 68\% to 9\% in the lung and from 26\% to $<1\%$ in the heart.
For a simple minipig plan, the full-width-half-maximum of the dose, dose rate, and LET distributions decreased by 30\%, 1.6\%, and 57\%, respectively, again with similar target dose coverage, thus reducing uncertainty in these quantities for preclinical studies.

\noindent \textbf{Conclusion:}
The inverse solution to IPO-IMPT demonstrated the capability to simultaneously modulate sub-spot proton energy and intensity distributions for clinical and preclinical studies.

\noindent \keywords{FLASH proton therapy, linear energy transfer (LET), Monte Carlo simulation, radiotherapy optimization, lung stereotactic body radiation therapy (SBRT)}

\section{Introduction}
\label{sec:Introduction}

Lung stereotactic body radiation therapy (SBRT) has revolutionized the treatment of peripheral lung cancer, providing excellent local control, very low rates of high grade toxicity, and cures for some early-stage patients \cite{timmerman2018}.
There are, however, subgroups of patients for whom SBRT has higher toxicity, particularly those with locally advanced tumors in ultra-central locations that abut the esophagus or mainstem bronchus \cite{lindberg2021}\cite{stam2017}.
Proton FLASH radiotherapy is a novel treatment modality that uses ultra-high dose rates (UHDR), with a dose rate threshold of 40-100 Gy/s \cite{kim2022}\cite{diffenderfer2022} and a dose threshold of 4-10 Gy \cite{bohlen2022}, to spare organs at risk (OARs) during treatment while maintaining tumor killing efficacy \cite{vozenin2019}\cite{montaygruel2017}\cite{jones2022}.
FLASH, which typically delivers a dose within a few milliseconds, is also advantageous from the point of view of motion mitigation, which is a known challenge in thoracic and upper gastrointestinal radiotherapy.

Clinical implementation of FLASH is currently challenging due to the difficulty of active energy modulation on a millisecond time scale.
An alternative to active energy modulation is to use a transmission beam, but this can lead to unacceptable risks to distal and peripheral OARs \cite{rothwell2022}\cite{underwood2018}.
Passive energy modulation is perhaps the most promising approach for conformal delivery of FLASH fields \cite{wei2022}\cite{ma2023}, but designing filters is difficult \cite{rliu2023}.
Nevertheless, proton FLASH therapy has been used in clinical trials \cite{mascia2023}\cite{daugherty2023}, including the first in-human prospective study of FLASH for patients with bone metastases, showing feasibility of delivering FLASH treatment.

In addition to the UHDR sparing effect, proton therapy planning must consider linear energy transfer (LET), a quantity related to radiation quality that can have a large impact on biological effectiveness \cite{jones2022}\cite{paganetti2019}\cite{grassberger2011}\cite{an2017}.
Lack of LET optimization for conformal FLASH compromises both clinical outcomes and the ability to interpret preclinical studies, as extra biological dose (XBD) attributable to high LET at the distal edge of the Bragg peak potentially offsets FLASH sparing \cite{le2006}\cite{friedl2022}.

To reduce the risk of fatal lung toxicity, single-fraction photon SBRT is subject to volume constraints at 5 and 20 Gy dose levels \cite{nicosia2019}.
Proton irradiation has been reported to cause extra lung toxicities compared to photon irradiation due to higher LET ($>2$ keV/$\mu$m) \cite{underwood2018}\cite{li2019}.
Vozenin et al have deemed lung sparing as the most promising FLASH application, increasing dose tolerance from 17 to 30 Gy in a mouse model \cite{vozenin2019b}.
This result was also confirmed by Bohlen et al \cite{bohlen2022}.
Both studies used a mouse lung and reported wide spreads in dose modifying factor (DMF) (0.5-0.8), dose threshold (1-10 Gy), and dose rate threshold (40-100 Gy/s) for FLASH sparing, which could possibly be due to unaccounted for XBD(LET) \cite{jones2022}.

Recently, it has been shown that it is possible to achieve simultaneous dose, dose rate, and LET optimization using a novel Integrated Physical Optimization Intensity-modulated Proton Therapy (IPO-IMPT) treatment planning framework in combination with patient-specific ridge filters \cite{rliu2023}.
Simultaneous inverse optimization of dose and dose rate only \cite{gao2020}, or dose and LET only \cite{an2017}, had previously been achieved, but the IPO-IMPT framework considers all three of these quantities.
Generation of fully optimized solutions is daunting due to the very large search space of solutions, specified in maps of range compensation, range modulation, and proton intensity, time-consuming Monte Carlo (MC) simulation of the corresponding dose and LET influence matrices, and computation of the non-convex composite objective function terms.
Furthermore, all of this must be evaluated repeatedly for iterative minimization, i.e., optimization, in that search space.
Our initial solution to IPO-IMPT was thus heuristic, forward, and limited to range modulation to improve dose rate.
The shortcomings of this solution include: (1) it only considered completely removing certain filter pins while leaving others unchanged and kept the beam compensation length fixed, and (2) it lacks an integrated inverse optimization from IPO-IMPT objective to optimize the filter geometry and spot map.
Here we describe a new inverse optimization approach, termed Simultaneous Intensity and Energy Modulation and Compensation (SIEMAC), that simultaneously specifies the weights of a proton intensity map and the geometry of a sparse filter, i.e. sparse energy modulation and compensation.

Advancing FLASH radiotherapy to its full potential will require preclinical studies to derive XBD models that endeavor to separate contributions from dose, dose rate, and LET, each of which has a non-discrete value per irradiation.
This could potentially be a long and costly experimental process requiring a large number of animal irradiations due to the complicated, entangled dependence of dose, dose rate, and LET on biological response.
Unlike clinical plan optimization that aims for best sparing of OARs while maintaining tumor dose coverage, preclinical plans should be optimized to reach minimum spreads of XBD distributions within the OAR target, allowing more efficient derivation of XBD models and therefore requiring fewer irradiations.
In other words, having control over the average values and spreads of dose, dose rate, and LET distributions will make, for example, determination of dose and dose rate thresholds of FLASH sparing easier.
Here we demonstrate that SIEMAC can also be used for such preclinical applications to solve integrated \textit{biological} optimization IMPT (IBO-IMPT) using XBD.

\section{Methods}
\label{sec:Methods}

We present a preliminary inverse solution to IPO-IMPT, SIEMAC, which consists of iteratively optimizing the geometry of patient-specific sets of range compensating bars and range modulating pins, and the weights of a proton pencil beam spot map, to deliver more desirable dose rate and LET distributions.
For clinical applications, SIEMAC increases dose rate and reduces LET to OARs while maintaining minimal dose to OARs and conformality to the clinical target volume (CTV) and beam specific planning target volumes (BSPTVs) \cite{lin2015b} when compared with more conventional techniques.
For preclinical applications, SIEMAC reduces the spreads of the dose, dose rate, and LET distributions per OAR irradiation to minimize the entanglement of variables that affect the convergence of XBD model derivation.
The approach uses distributed parallel computing to make highly computationally intensive calculations manageable.
Varian Probeam specifications, including 300 nA maximum current at 250 MeV energy, were used in this study.

\subsection{Extending the Traditional IMPT Optimization Problem to Solve IPO-IMPT}
\label{subsec:IMPTExtended}

\begin{figure}[H]
\centering
\includegraphics[width=6.5in, trim={0 2px 0 0}, clip]{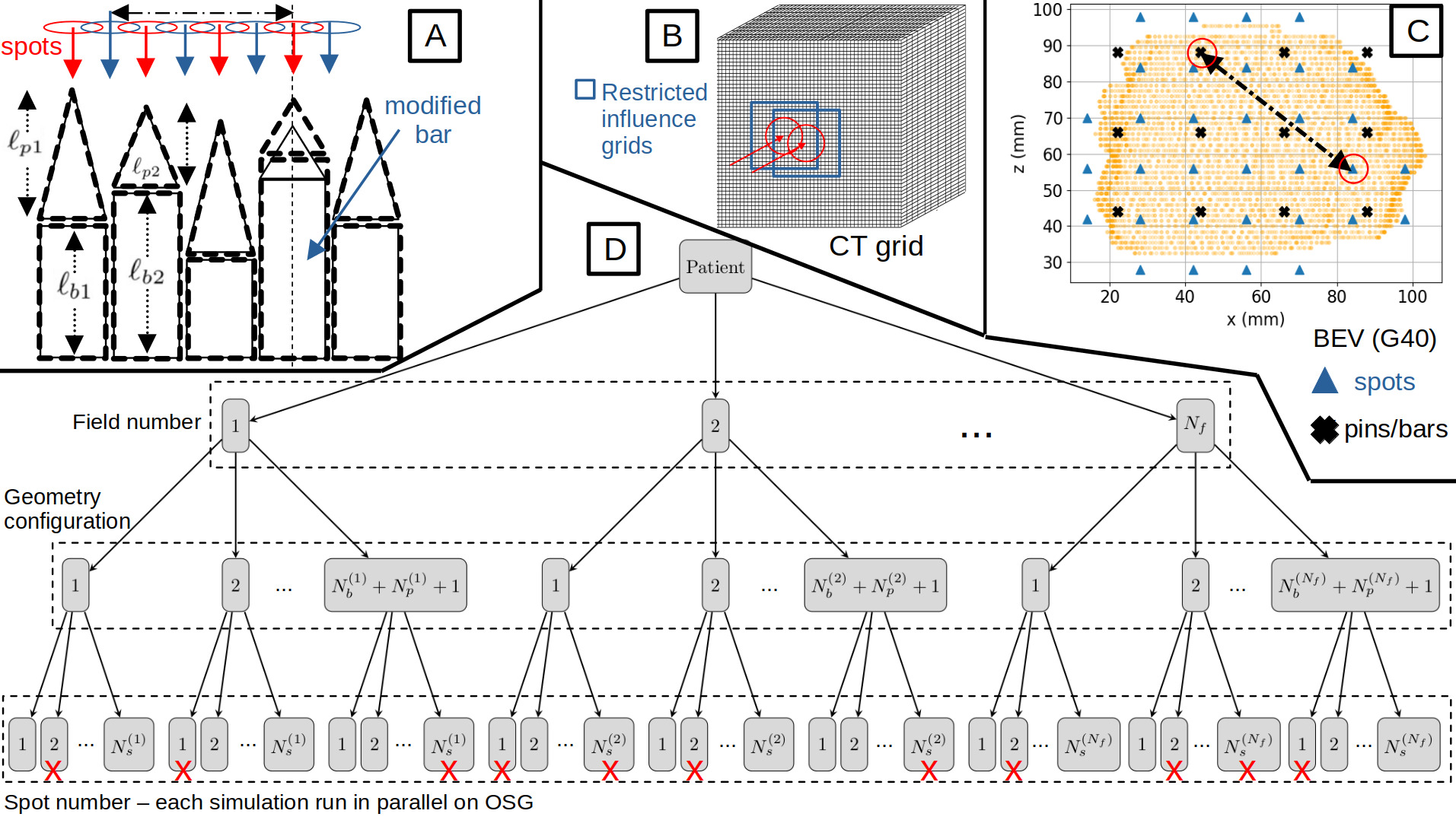}
\caption{
A: The beam's eye view (BEV) for a specific field of the CTV (small orange points), spot map (blue triangles), and pin/bar locations (black X symbols) (spacing increased for visual clarity). The dot-dash line with double-ended arrows represents the distance between a particular spot and the geometry component.
B: A 3D grid representing the voxelized patient CT. Each spot (red arrow/circle) has a unique restricted influence grid (blue box) that is much smaller than the full CT grid.
C: Side view of 5 pins, 5 bars, and 8 spots (colorized for visual clarity) showing the interjoining of spots and pins. The length of each pin or bar is variable. The dot-dash line with double-ended arrows represents the distance between a particular spot and the geometry component.
D: Simulation parallelization scheme for one patient and $N_f$ fields. A red X in the bottom row represents a simulation that can be skipped due to the spot being far away from the modified geometry component (represented by the dot-dash line with double-ended arrows in panels A and C).
}
\label{fig:method_figure}
\end{figure}

Traditional IMPT optimizes the weights ($\dub$) of a pencil beam spot map in order to produce a conformal dose distribution.
A brief summary of traditional IMPT optimization can be found in Appendix \ref{sec:TraditionalIMPT}, where important quantities such as the dose influence matrix ($D_{ij}$), objective function ($f$), prescribed dose ($\hat{d}$), and penalty factor ($p$) are also defined.
In this work, we expand the arguments of the objective function to include geometry parameters.
Specifically, $\lb \in \Real^{N_b}$ are the lengths of the range compensating bars (bars for short) with $N_b$ representing the number of bars,
and $\lp \in \Real^{N_p}$ are the lengths of the range modulating pins (pins for short) with $N_p$ representing the number of pins, as shown in Figure~\ref{fig:method_figure}A.
Usually, $N_b = N_p$, but this is not strictly necessary, so they are kept as two separate variables.
A summary of these geometry components are in Figure~\ref{fig:terminology}.
Figure~\ref{fig:method_figure}A also shows the interjoining of spots and pins, i.e. the red spots impinge on pin peaks and the blue spots impinge on the valleys.
Furthermore, since IMPT is typically delivered using multiple fields, we use superscripts on the variables to identify to which field that variable belongs (e.g. $N_b^{(2)}$ is the number of bars for field number 2) and use $N_f$ to represent the total number of fields.

Previous work in our laboratory and elsewhere used ziggurat-shaped pins \cite{simeonov2017}\cite{rliu2023} to create spread-out Bragg peaks (SOBPs).
Here, we adopt a simpler square pyramid-shaped pin that accomplishes the same task, but requires less computational effort to design.

Additionally, we also expand the objective function to include dose rate and LET objectives.
Thus, the new problem to solve becomes
\begin{equation}
\label{eq:newimpt1}
\substack{\text{argmin} \\ \dub, \lb, \lp} f \left( \dub, \lb, \lp \right)
\end{equation}
where
\begin{equation}
\label{eq:newimpt2}
f \left( \dub, \lb, \lp \right) = \sum\limits_n p_n f_n^D \left( \dub, \lb, \lp \right) + \sum\limits_n p_n f_n^{DR} \left( \dub, \lb, \lp \right) + \sum\limits_n p_n f_n^{LET} \left( \dub, \lb, \lp \right).
\end{equation}
The dose rate and LET objectives, $f_n^{DR}$ and $f_n^{LET}$, can be easily defined in a way directly analogous to Equation~\ref{eq:sqdev}, and the arguments are again typically constrained by upper and lower bounds.
More specifically, the objective function used in this analysis is
\begin{equation}
\label{eq:newimpt_specific}
\begin{split}
f \left( \dub, \lb, \lp \right) = & \sum\limits_{BSPTV} \frac{p_{BSPTV}}{N_{BSPTV}} \sum\limits_{i \in BSPTV} \Theta \left( \frac{\hat{d}}{N_f} - d_i \right) \left( d_i - \frac{\hat{d}}{N_f} \right)^2 \\
& + \frac{p_{CTV}}{N_{CTV}} \sum\limits_{i \in CTV} \Theta \left( \hat{d} - d_i \right) \left( d_i - \hat{d} \right)^2 \\
& + \sum\limits_{OAR} \frac{p_{OAR}}{N_{OAR}} \sum\limits_{\substack{i \in OAR \\ d_i > D_0}} \Theta \left( \widehat{DR} - DR_i \right) \left( DR_i - \widehat{DR} \right)^2 \\
& + \sum\limits_{OAR} \frac{p_{OAR}}{N_{OAR}} \sum\limits_{\substack{i \in OAR \\ d_i > D_0}} \Theta \left( LET_i - \widehat{LET} \right) \left( LET_i - \widehat{LET} \right)^2 \\
& + \frac{p_{ROB}}{N_{ROB}} \sum\limits_{i \in ROB}  d_i^2
\end{split}
\end{equation}
subject to upper and lower bounds on each optimization variable:
\begin{equation}
\label{eq:bounds}
v_{i, min} < v_i < v_{i, max}
\end{equation}
where the generic variable $\boldsymbol{v}$ has been introduced for simplicity to represent the concatenation of $\dub$, $\lb$, and $\lp$; and $\hat{d}$, $\widehat{DR}$, and $\widehat{LET}$ are the prescription dose, target dose rate, and target LET, respectively; and $\Theta$ is the Heaviside function.
$D_0$ is a dose cutoff, where voxels with a dose below this value are not considered in the objective; typical values are 5\% - 10\% of the prescribed dose \cite{folkerts2020}.
ROB refers to the rest-of-body which is everything in the body besides the CTV and BSPTVs.
Since dose rate and LET have contributions from each spot, we use dose averaged dose rate \cite{vandewater2019} and LET, \textit{i.e.}
\begin{equation}
\label{eq:DADR}
DR_i = \frac{ \sum\limits_j D_{ij}^2 w_j I_j }{ q_e \sum\limits_j D_{ij} w_j }
\end{equation}
and
\begin{equation}
\label{eq:LETd}
LET_i = \frac{ \sum\limits_j LET_{ij} D_{ij} w_{j} }{ \sum\limits_j D_{ij} w_j }
\end{equation}
where $DR_i$ is the DADR in voxel $i$, $I_j$ is the nozzle current of spot $j$ (300 nA in this work), $LET_i$ is the dose averaged LET in voxel $i$, and $LET_{ij}$ (the LET influence matrix) is the dose averaged LET in voxel $i$ due to spot $j$.

Here we introduce the concept of a restricted influence grid (RIG) to limit the extent of dose and LET influence matrices by inclusion of the spots that are within FLASH millisecond timing proximity of the location of the highest instantaneous dose.
A RIG exists for each voxel $i$, and consists of voxel $i$ plus the neighboring voxels surrounding it.
A time value for each RIG can then be defined as
\begin{equation}
\label{eq:tRIGi}
t^{RIG}_i = \frac{q_e \sum\limits_j w_j f_{ij}}{I} = \sum\limits_j f_{ij} t_j^{spot}
\end{equation}
where $f_{ij}$ is the fraction of spot $j$ that impinges on RIG $i$ and $t^{spot}_j = q_e w_j/I$ is the actual time duration of spot $j$.
Alternatively, $f_{ij}$ could also be defined as a Boolean value equal to 1 when the threshold of 0.5 is met, and 0 otherwise.
In other words, $f_{ij} t_j^{spot}$ is the hypothetical irradiation time on RIG $i$ from spot $j$ assuming the entire spot impinges on RIG $i$ rather than just a fraction of it, and $t^{RIG}_i$ is a sum over these values without accounting for scan time + delivery time of other spots.
Thus, $t_{ij}^{RIG} \approx t_j^{spot}$ when the spot and RIG mostly overlap, and $t_{ij}^{RIG} << t_j^{spot}$ when the spot and RIG overlap very little.
This concept of most of the dose in a give voxel being delivered within a relatively short time window (compared to the full irradiation time) was introduced in \cite{folkerts2020}.

For simplicity, we have adopted a very rudimentary version of a RIG for this work and use a restricted dose (or LET) influence matrix:
\begin{equation}
\label{eq:Dij_restricted}
D_{ij} =
\begin{cases} 
  D_{ij}^{\text{unrestricted}} & i \in \text{spot specific restricted influence grid} \\
  0 & \text{otherwise}
\end{cases}
\end{equation}
where $D_{ij}^{\text{unrestricted}}$ is the dose to voxel $i$ due to spot $j$ considering the entirety of the CT grid.
The restricted $D_{ij}$, illustrated in Figures~\ref{fig:method_figure}B and \ref{fig:sim_with_restricted_grid}, significantly trims down the CT grid for the sake of computational performance by assuming the dose is negligible in voxels far away from the spot.
The dot-dash line with double-ended arrows in Figures~\ref{fig:method_figure}A and \ref{fig:method_figure}C show the spots interjoining with sparse pins/bars subject to RIG.

These changes to the optimization problem present several challenges:
\begin{enumerate}
  \item The added arguments and objectives make the problem more complex and make solving the problem more CPU intensive. Furthermore, $L_{ij}$ must also be calculated in addition to $D_{ij}$.
  \item The added arguments and objectives can make the problem non-convex.
  \item The variability of the geometry parameters means that $D_{ij}$ and $L_{ij}$ need to be re-calculated many times as the geometry changes, and it also makes the gradient calculation much more difficult since $D_{ij}$ and $L_{ij}$ are not constant.
  This leads to a further and very significant increase in necessary computing power.
\end{enumerate}

\subsection{Approach to Solving IPO-IMPT}
\label{subsec:OptimizationApproach}

To address these challenges, a parallel computing framework has been developed, and we use anonymized patient data to demonstrate the technique.

The first step is to define initial (i.e. zeroth order) geometry.
This can be done in several ways:
\begin{enumerate}
  \item Using ray tracing to design a filter meant to produce a conformal dose distribution \cite{simeonov2017}.
  \item Using a forward heuristic such as the sparse modulation technique described by \cite{rliu2023}.
  \item Using a global search algorithm such as differential evolution \cite{storn1997} or dual annealing \cite{xiang1997}.
\end{enumerate}
This is an important step because the goal of the remaining steps is to find a local minimum of the objective function in the neighborhood of this starting point,
so it is important that the starting point is within a desirable valley.
There is no straightforward way of deciding how to choose this starting point, and several different starting points may need to be tried.

Next, we simply use a quasi-Newton method (the L-BFGS-B algorithm) to better optimize the initial geometry, along with the spot weights.
The challenge here, however, is that this is extremely CPU intensive and requires a powerful computing cluster to succeed.
We used the Open Science Grid (OSG) \cite{pordes2007}\cite{sfiligoi2009}\cite{osg2006} and performed the simulations with TOPAS \cite{faddegon2020}.

The gradient of the objective function is
\begin{equation}
\label{eq:gradf}
\nabla f = \left(
\frac{\partial f}{\partial w_1} , ... , \frac{\partial f}{\partial w_{N_s}} ,
\frac{\partial f}{\partial \ell_{b1}} , ... , \frac{\partial f}{\partial \ell_{bN_b}} ,
\frac{\partial f}{\partial \ell_{p1}} , ... , \frac{\partial f}{\partial \ell_{pN_p}}
\right).
\end{equation}
The partial derivatives $\frac{\partial f}{\partial w_1} , ... , \frac{\partial f}{\partial w_{N_s}}$ are straightforward to calculate analytically.
The remaining partial derivatives are estimated using the finite difference approximation
\begin{equation}
\label{eq:finitediff}
\frac{\partial f}{\partial v_i} = \frac{ f \left( v_1, ..., v_i + \Delta v , ... , v_n \right) - f \left( \boldsymbol{v} \right) }{\Delta v}.
\end{equation}
Since $f$ depends on $D_{ij}$ and $L_{ij}$, and since $D_{ij}$ and $L_{ij}$ depend on the pin and bar lengths,
it can be seen in Equations~\ref{eq:gradf} and \ref{eq:finitediff} that the number of geometries, and therefore the number of $D_{ij}$'s and $L_{ij}$'s that need to be calculated with MC, is $N_b + N_p + 1$ for each field and for each iteration of the optimization.

In order to complete these calculations in a reasonable amount of time, they were broken down into parallelizable pieces and submitted to OSG.
The exact parallelization scheme is illustrated in Figure~\ref{fig:method_figure}D.
A red X in the bottom row of Figure~\ref{fig:method_figure}D represents a simulation that can be skipped due to the spot being far away from the modified geometry component (represented by the dot-dash line with double-ended arrows in Figures~\ref{fig:method_figure}A and C), which therefore saves time.

\begin{figure}[H]
\centering
\includegraphics[width=6.5in]{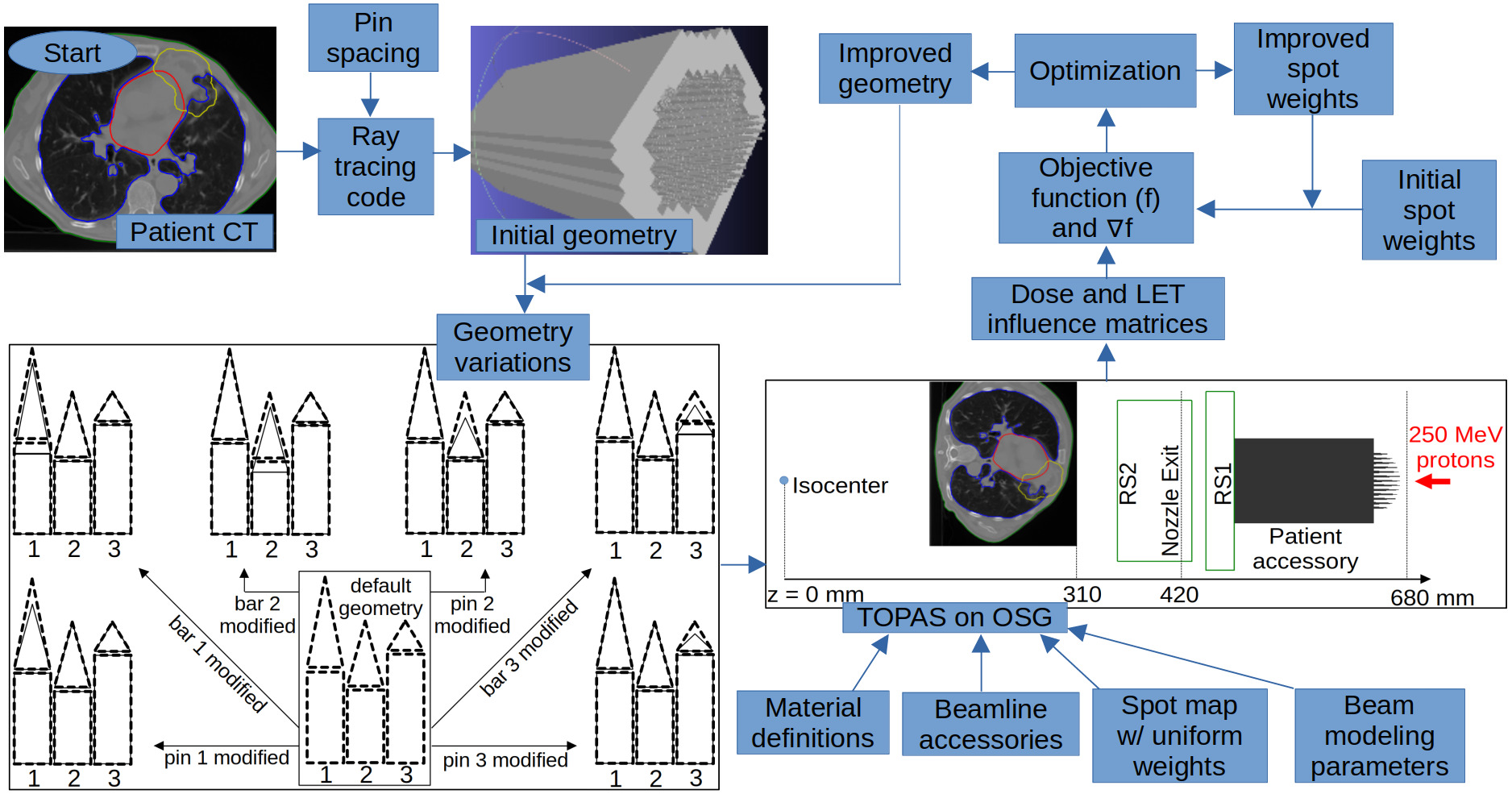}
\caption{Workflow that employs TOPAS MC to iteratively optimize the geometry of patient specific sets of pins and bars and spot maps for dose, dose rate, and LET.}
\label{fig:workflow}
\end{figure}

The overall workflow of the optimization can be seen in Figure~\ref{fig:workflow}.
The process begins by using a ray tracing algorithm with a patient CT to define the initial geometry of the pins and bars.
A TOPAS MC simulation is then used to calculate $D_{ij}$ and $L_{ij}$.
In parallel, many geometry variations are also simulated that are needed to calculate the gradient of the objective function.
The simulation output data are then fed into an optimization algorithm and the process is repeated until an acceptable solution is reached.

\subsection{IPO-IMPT for preclinical study}
\label{subsec:AnimalStudies}

Unlike clinical treatment plans, preclinical studies typically aim to irradiate an OAR, and therefore require different optimization objectives.
Preclinical objectives should include minimizing the spreads of the dose, dose rate, and LET distributions in the OAR target, thereby minimizing uncertainty when separating the contributions from each of these quantities on XBD.
The SIEMAC algorithm was tested to see if it is feasible to indirectly optimize XBD via the physical quantities of dose, dose rate, and LET.
The objective function, Equation~\ref{eq:animal_objective}, was first set to deliver a uniform dose of 20 Gy to the target, which represents the threshold for short-term pneumonitis and long-term fibrosis \cite{le2006}\cite{nicosia2019}\cite{friedl2022}.
In other words, the last two lines of Equation~\ref{eq:animal_objective} were not used initially.
\begin{equation}
\label{eq:animal_objective}
\begin{split}
f \left( \dub, \lb, \lp \right) = & \frac{p_{TARG}}{N_{TARG}} \sum\limits_{i \in TARG} \left( d_i - \hat{d} \right)^2 \\
& + \frac{p_{ROB}}{N_{ROB}} \sum\limits_{i \in ROB}  d_i^2 \\
& + \frac{p_{TARG}}{N_{TARG}} \sum\limits_{i \in TARG} \left( DR_i - \widehat{DR} \right)^2 \\
& + \frac{p_{TARG}}{N_{TARG}} \sum\limits_{i \in TARG} \left( LET_i - \widehat{LET} \right)^2
\end{split}
\end{equation}
Then, a second round of optimization was done that included the last two lines of Equation~\ref{eq:animal_objective} to attempt to narrow the dose, DADR, and $\LETd$ distributions, and therefore reduce the uncertainty in these quantities, while maintaining similar target dose coverage.
We compared the magnitude of spreads of dose, dose rate and LET distributions and their $XBD_i (DADR)$ and $XBD_i (LET)$ on a 36-mm spherical irradiation target of a minipig lung before and after IPO-IMPT.

XBD comes in two different flavors, $XBD_i (DADR)$ and $XBD_i (LET)$ (where $i$ is the voxel number), defined as
\begin{equation}
\label{eq:xbddr}
XBD_i (DADR) = d_i \frac{k}{1 + e^{-a \left( DR_i - DR_t \right)}}
\end{equation}
and
\begin{equation}
\label{eq:xbdlet}
XBD_i (LET) = d_i c LET_i
\end{equation}
which represent adjustments to the physical dose that take into account biological responses to radiation.
Advantages to healthy tissue are represented by larger values of $XBD_i (DADR)$ and smaller values of $XBD_i (LET)$.
Here, $a$, $k$, $DR_t$, and $c$ are parameters that depend on biological mechanisms.

\begin{figure}[H]
\centering
\includegraphics[width=4.0in]{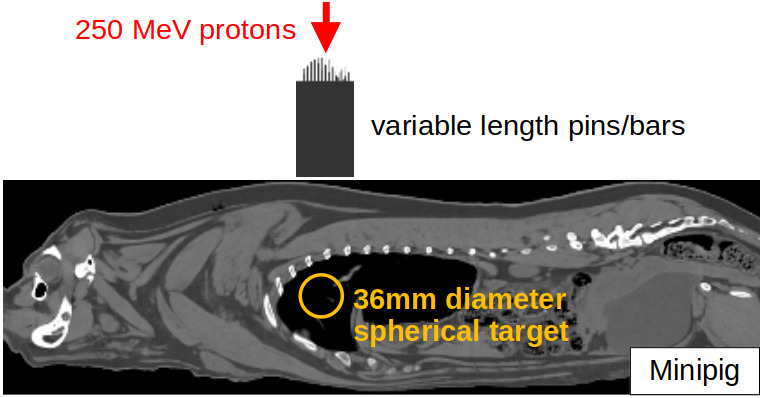}
\caption{
Design for the minipig study consisting of an anterior 250 MeV proton pencil beam and sets of variable length pins and bars that can be optimized to irradiate a 36 mm diameter sphere within the lung.
}
\label{fig:animal_study_method}
\end{figure}

The design of the minipig simulations is shown in Figure~\ref{fig:animal_study_method}, which includes an anterior 250 MeV proton pencil beam and sets of variable length pins and bars that can be optimized to irradiate the spherical target.

\section{Results}
\label{sec:Results}

\subsection{Lung Cancer Patient Plan}
\label{subsec:PatientResults}

\begin{figure}[H]
\centering
\includegraphics[width=6.5in]{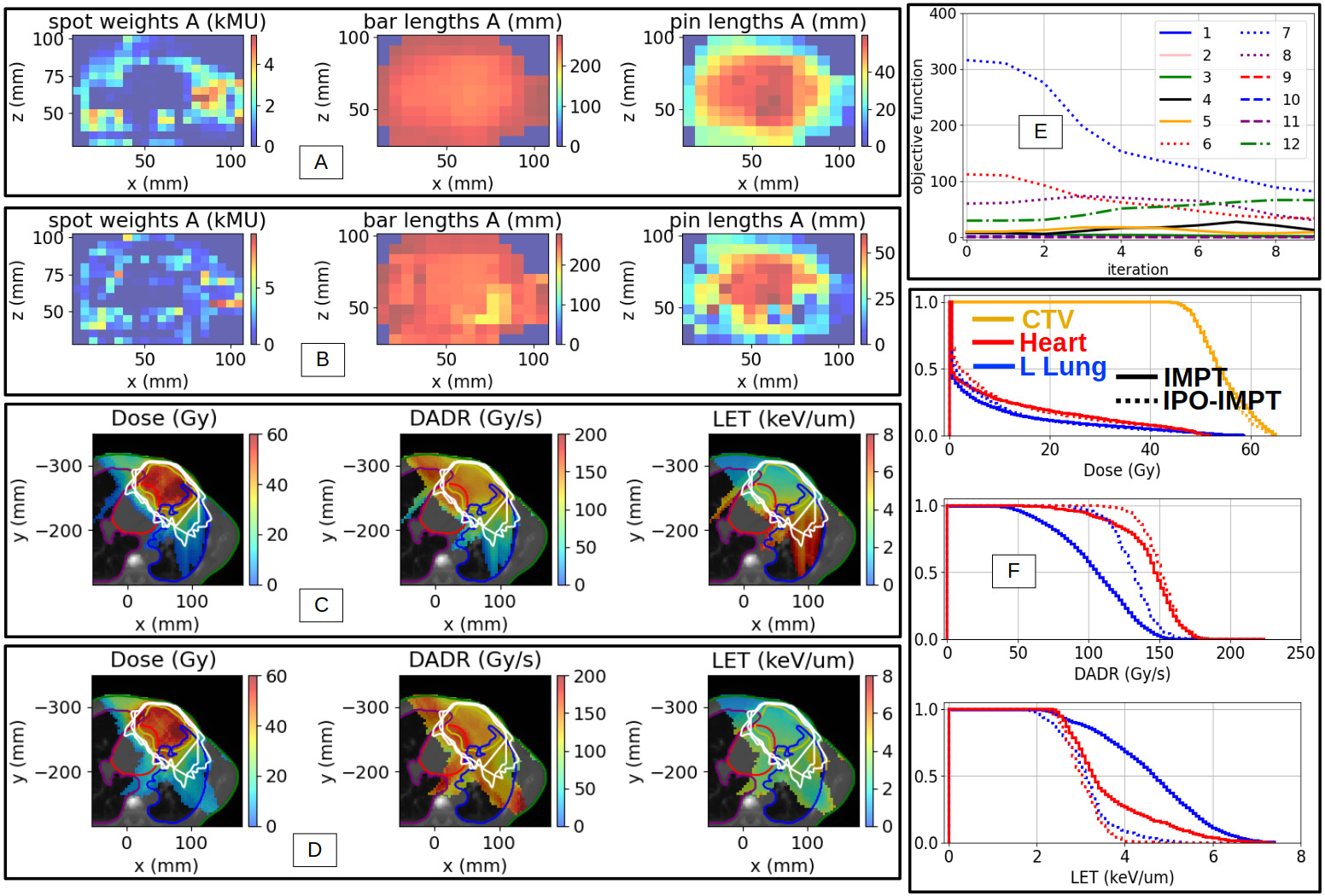}
\caption{IPO-IMPT optimization results.
A and B: Maps of spot weights, bar lengths (i.e. sparse compensation), and pin lengths (i.e. sparse modulation) for field A (gantry 40) before and after IPO-IMPT optimization, respectively.
C and D: Dose, dose rate, and LET distributions for an axial slice of the 3-field patient plan before and after IPO-IMPT optimization, respectively.
Also shown are contours of the left lung (blue), right lung (purple), heart (red), CTV (yellow), and three BSPTVs (white).
E: The 12 components of the objective function (Equation~\ref{eq:newimpt_specific}) as a function of iteration number.
F: Dose (top), dose rate (middle), and LET (bottom) volume histograms for the CTV (yellow), left lung (blue), and heart (red) before (solid line) and after (dashed line) IPO-IMPT optimization.
Dose is reported for the whole structures (CTV, lung, heart), while DADR and $\LETd$ are reported only for the evaluation structures, as in our previous work \cite{rliu2023}.
}
\label{fig:result}
\end{figure}

To demonstrate SIEMAC, we created a three-field treatment plan for a representative lung cancer patient.
The dose prescription to the centrally located CTV was 50 Gy, with nearby OARs consisting of heart and left lung.
Figure~\ref{fig:result} summarizes the result.
Panel A shows the spot map, bar lengths, and pin lengths for one of the three fields used (field A = gantry 40, field B = gantry 0, field C = gantry 320) in this study for iteration 0.
We define iteration 0 to be the result after spot-weight-only optimization (i.e., the geometry parameters are held fixed) has been done using traditional IMPT techniques.
Panel B is the same as panel A except after 9 iterations.
Similarly, panels C and D show before and after distributions of dose, dose rate, and LET for an axial slice of the patient.
Panel E shows the different components of the objective function vs the optimization iteration number.
Finally, panel F shows dose, dose rate, and LET volume histograms for the CTV, left lung, and heart.

The plots in Figure~\ref{fig:result}F show sizeable improvements to the dose rate and LET distributions in the lung and heart, with a negligible sacrifice to the dose distributions, when comparing traditional IMPT to IPO-IMPT with SIEMAC.
For the OARs, we use an evaluation volume, which refers to the overlap between the OAR and BSPTV, excluding the CTV and any voxels with dose below 4 Gy, as in our previous work \cite{rliu2023}.
For the heart and lung, the percentage of the evaluation volume receiving above the FLASH threshold of 100 Gy/s rose from 93\% to 100\% and from 57\% to 96\%, respectively.
Additionally, $\LETd$ coverage above 4 keV/um dropped from 68\% to 9\% in the lung and from 26\% to $<1\%$ in the heart.
These improvements can be attributed to the shortening of up to 100 mm bar length and 60 mm pin length (Figure~\ref{fig:result}A-B), demonstrating improvements to sparse compensation and sparse modulation, along with improved spot maps that were optimized simultaneously with the pin and bar lengths.
The sparse compensation in particular might explain the improvements in the LET distribution over our initial forward heuristic solution.

\subsection{Preclinical optimization}
\label{subsec:AnimalResults}

To demonstrate this functionality of SIEMAC, we have generated a single field plan for an animal irradiation with narrower dose, dose rate, LET, and XBD distributions in the irradiated OAR compared to an unoptimized plan, therefore reducing the uncertainty in these variables when deriving XBD models from preclinical studies.
Values of $c = 0.04 \ \mu m / keV$, $DR_t = 40 \ Gy/s$, $k = 0.5$, and $a = 4/DR_t$, were used in equations \ref{eq:xbddr} and \ref{eq:xbdlet} for this work.
A plot of equation \ref{eq:xbddr} with these values is shown in Figure~\ref{fig:xbd_plot}, which shows a partial advantage and steepest gradient at 40 Gy/s, and a full advantage at 100 Gy/s.

\begin{figure}[H]
\centering
\includegraphics[width=6.5in]{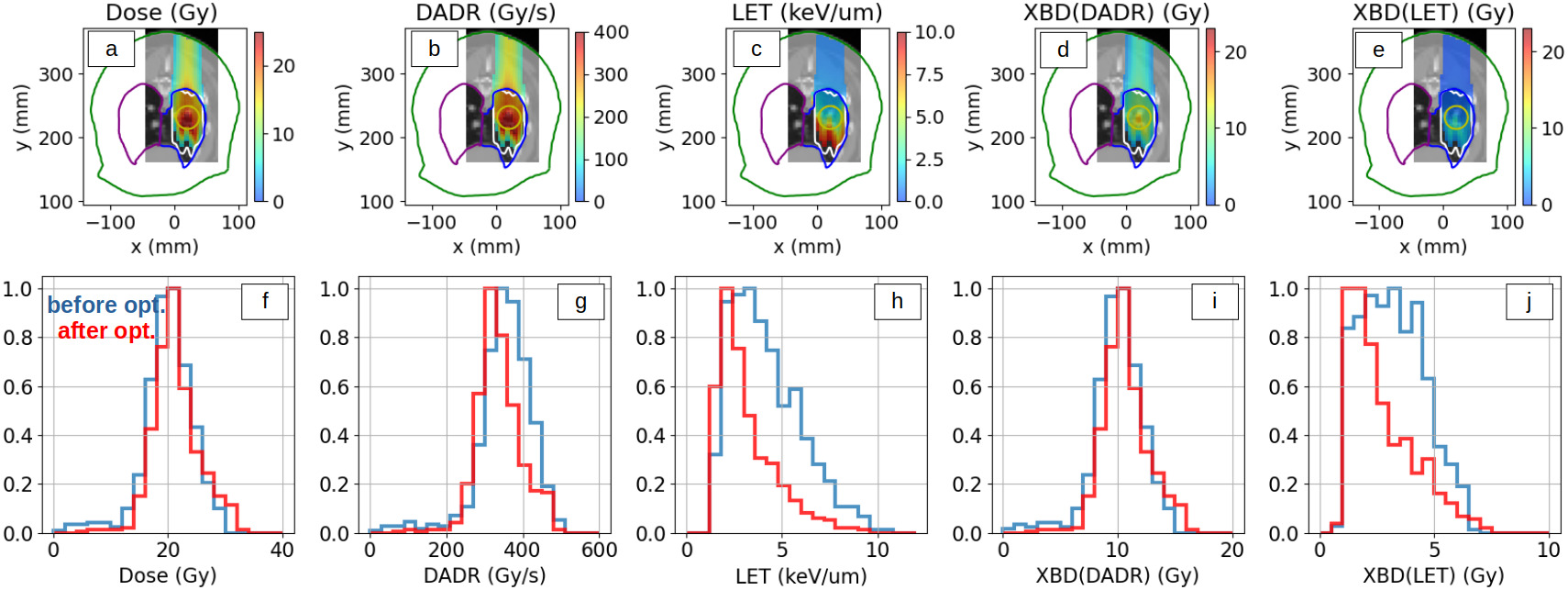}
\caption{Animal study results showing dose (a and f), dose rate (b and g), LET (c and h), XBD(DADR) (d and i), and XBD(LET) (e and j) (non-cumulative) distributions for the spherical target of the minipig for the design described in described in Figure~\ref{fig:animal_study_method}.
The bottom row (f-j) shows results before (blue) and after (red) optimization was done to reduce the spread in dose, DADR, and $\LETd$.}
\label{fig:animal_results}
\end{figure}

Figure~\ref{fig:animal_results} summarizes the results of the SIEMAC generated plan, which shows dose, dose rate, LET, XBD(DADR), and XBD(LET) distributions for the target in the left lung of the minipig before and after optimization.
We quantify the improvements to the optimized plan by reporting both the full-width-half-maximum (FWHM) of each distribution, which is large for undesirable widely spread distributions and approaches zero for ideal distributions, as well as the area under the histograms after normalizing to a maximum value of 1, which, similarly, is large for undesirable distributions and smaller for ideal distributions.
The results show that SIEMAC can be used to reduce the unoptimized wide spread in dose, DADR, and $\LETd$ (red vs blue lines in Figure~\ref{fig:animal_results} panels f-j) distributions in animal studies.

For dose, SIEMAC decreased the FWHM by 30\% (10 Gy to 7 Gy) and the area of the normalized histogram by 15\% (4.8 to 4.1 a.u.).
For DADR, the FWHM decreased by 1.2\% (122 Gy/s to 120 Gy/s) and area decreased by 21\% (4.8 to 3.8 a.u.).
And for LET, FWHM decreased by 57\% (7.1 keV/$\mu$m to 4.0 keV/$\mu$m) and area decreased by 44\% (7.1 to 4.0 a.u.).
To associate extra toxicity (i.e. biological effect) due to dose rate and LET distributions, XBD(DADR) and XBD(LET) are calculated using the proposed XBD model described in equations \ref{eq:xbddr} and \ref{eq:xbdlet}.
The inverse solution of IPO-IMPT demonstrated a modest reduction of XBD(DADR) because the optimization algorithm considers the unoptimized DADR is well above the full UHDR benefit of 100 Gy/s (Figure~\ref{fig:animal_results}g), at 300 nA nozzle current.
Such inverse solution of IBO-IMPT can improve much more XBD(DADR) for other organs and other beam conditions when needed as demonstrated for XBD(LET) (Figure~\ref{fig:animal_results}h).
In summary, the results shows a sizable XBD(LET) with a wide FWHM and area without optimization therefore XBD(LET) must be considered and optimzed when studying UHDR sparing of lung toxicity.

\section{Discussion}
\label{sec:Discussion}

These results demonstrate a proof-of-concept that SIEMAC can be used to produce proton FLASH treatment plans that provide considerable improvements over existing planning algorithms (Figure~\ref{fig:result} for clinical results and Figure~\ref{fig:animal_results} for preclinical results).
Our inverse SIEMAC solution improves upon our initial forward heuristic solution by iteratively optimizing range modulation, range compensation, and spot intensity map.
The solution providng an opportunity to modulate sub-spot proton energy and proton intensity, which are vital for microscale radiation transport and thus FLASH optimization for simultaneous improvements in dose rate and LET of OARs.
The technique has been shown to also be useful in animal studies for narrowing the dose, dose rate, and LET distributions in the target, therefore making derivation of XBD models more efficient and less uncertain.

MC simulation \cite{faddegon2020}\cite{schuemann2019} of radiation transport and biochemical processes in microscale timing and spatial dimensions for each of the $\mathcal{O}(10^9)$ incoming protons is too time consuming, even for super computers, given the complexity of quantum physics equations embedded and implemented by MC.
However, Schuemann et al. have simplified the full, LET dependent, quantum physics processes in complex MC at microscale radiation transport to simulate the FLASH biological effects \cite{schuemann2019}\cite{ramosmendez2020} and recommended that LET be included in FLASH optimization \cite{taylor2022}.
This manuscript responds to this challenge by providing an inverse solution to IPO-IMPT that can be implemented on a modest cluster.
Such an inverse solution to IPO-IMPT can potentially improve cancer patient outcomes because microscale radiation transport underlies biochemical processes responsible for FLASH sparing of OARs.

The optimization technique described in this work is flexible enough that additional optimization parameters and objectives may be easily added.
For example, we found that the downstream distance from the nozzle of the patient can have a significant impact on dose rate.
We chose to keep this distance fixed, but this value could be made variable and included in the optimization.
Similarly, other quantities such as material density, beam current, etc., were kept fixed but could theoretically be optimized using our technique.
This may be the subject of future work, and the recently released TopasOpt \cite{whelan2023} may facilitate such studies.

Future work may also include fine-tuning the objective function to account for the varying radiosensitivities of different OARs.
For example, Brodin and Tom\'e report suggested dose constraints that vary by more than a factor of 2 across different OARs.
Esophagus and central airways, the central serial OARs, are typically easier to spare as shown by our initial forward heuristic solution of IPO-IMPT \cite{rliu2023}.
These serial OARs are therefore not included here for the sake of simplicity.
Additional considerations, such as the variability of proton relative biological effectiveness (RBE) across a multitude of physical and biological variables described by McNamara et al. \cite{mcnamara2020} and Peeler et al. \cite{peeler2016}, may be taken into account as well.

In our experience with this analysis, we were able to run up to about 3000 parallel jobs at once on OSG (while any additional jobs idled until a slot opened up), and each iteration took several hours.
While this is a significant advantage over a desktop computer, it does not represent peak performance that is achievable with dedicated modern supercomputers.
Therefore, additional performance improvements are still available using such resources.
Furthermore, we believe that with more time and resources, the efficiency of our code and workflow could be improved as well.
Nonetheless, the present studies provide valuable proof of concept and significantly advance FLASH treatment planning.

A possible concern is that dose rate and LET improvements came at the expense of more dose to the ROB.
Indeed, it will be important to quantify and explore these tradeoffs further.
Given the high degeneracy of solutions, such tradeoffs can be managed by making adjustments to the objective function.

In our limited preclinical feasibility study, we focused on the reduction of the unoptimized wide spread of dose, XBD(DADR) and XBD($\LETd$) distributions using IPO-IMPT for a minipig lung to help with quick convergence of XBD model derivations.
The preliminary case was chosen to show the capability of IPO-IMPT to inversely solve the most relevant issues for preclinical application, because lung is considered to have largest benefit from FLASH sparing with the biggest impact for the most fatal ultra-central lung cancer.
Inverse solution to IBO-IMPT and other organs, such as the esophagus, central airways, and heart, will be studied.
Figure~\ref{fig:animal_results} shows that it is feasible for our preliminary SIEMAC to reduce the spreads of dose, dose rate, and LET distributions.
Giving researchers control over the average values and spreads of dose, dose rate, and LET distributions can minimize the overlaps of dose, dose rate, and LET among irradiations, and therefore improve the efficiency with which XBD models can be derived and reduce the number of needed animal irradiations.
Alternative methods of quantifying the distribution spreads, besides FWHM and integrated area, could also prove to be more useful and will require additional studies.
Minimization of overlaps of dose, dose rate, and LET among irradiations are vital for biologists to separate their XBD(DADR) and XBD($\LETd$) from physical dose contribution, which can be badly entangled among these three terms and multiple irradiations without careful optimization, in observed OAR toxicities.

In addition to XBD, alternative definitions of dose rate besides DADR may prove to be more useful \cite{folkerts2020}, and could allow more elegant XBD(DADR) and XBD($\LETd$) models, and will also require further investigation.
Our definition of $t^{RIG}_i$ (equation \ref{eq:tRIGi}) also offers future opportunities for improvement.
Associated with spot peak dose rate \cite{vanmarlen2020}, RIG can potentially provide solutions more relevant to FLASH biology.
Although SIEMAC's current tapping of the underlying quantum physics radiation transport is rudimentary, further accessing the three microscopic dimensions and micro-timing is possible with better computing power and better implementations of inverse optimization.

\section{Conclusions}
\label{sec:Conclusions}

We have developed a novel iterative inverse optimization approach that uses parallel distributed computing to solve IPO-IMPT.
The optimization facilitates simultaneous solutions of range compensator, range modulator, and proton intensity modulation distributions, leveraging the underlying quantum physics processes of proton-tissue interactions to better optimize dose rate and LET.
The inverse optimization works for humans to increase dose rate and reduce LET in OARs.
With modifications to the objective function, it can also be used for preclinical studies to decrease uncertainty in dose, dose rate, and LET in derivations of XBD models by giving researches more control over these quantities.

\section{Acknowledgements}
\label{sec:Acknowledgements}

This research was done using services provided by the OSG Consortium \cite{pordes2007}\cite{sfiligoi2009}, which is supported by the National Science Foundation awards \#2030508 and \#1836650.
Additional funding was provided by Dr. Liyong Lin's Emory faculty fund.
We would also like to thank Emory dosimetrists and physicists for help with contouring.

\section{Supplemental}
\label{sec:Supplemental}

\renewcommand{\thefigure}{S\arabic{figure}}
\setcounter{figure}{0}

\begin{figure}[H]
\centering
\includegraphics[width=6.5in]{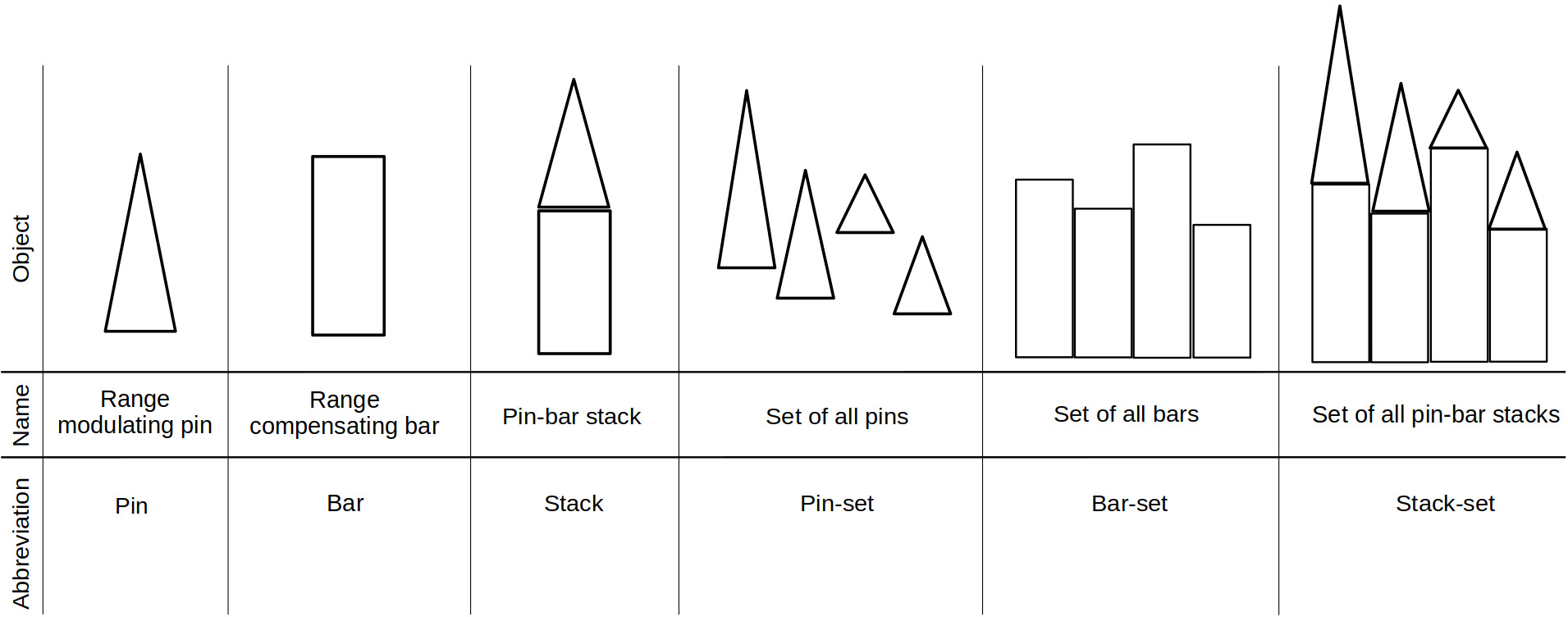}
\caption{Geometry component terminology.}
\label{fig:terminology}
\end{figure}

\begin{figure}[H]
\centering
\includegraphics[width=6.5in]{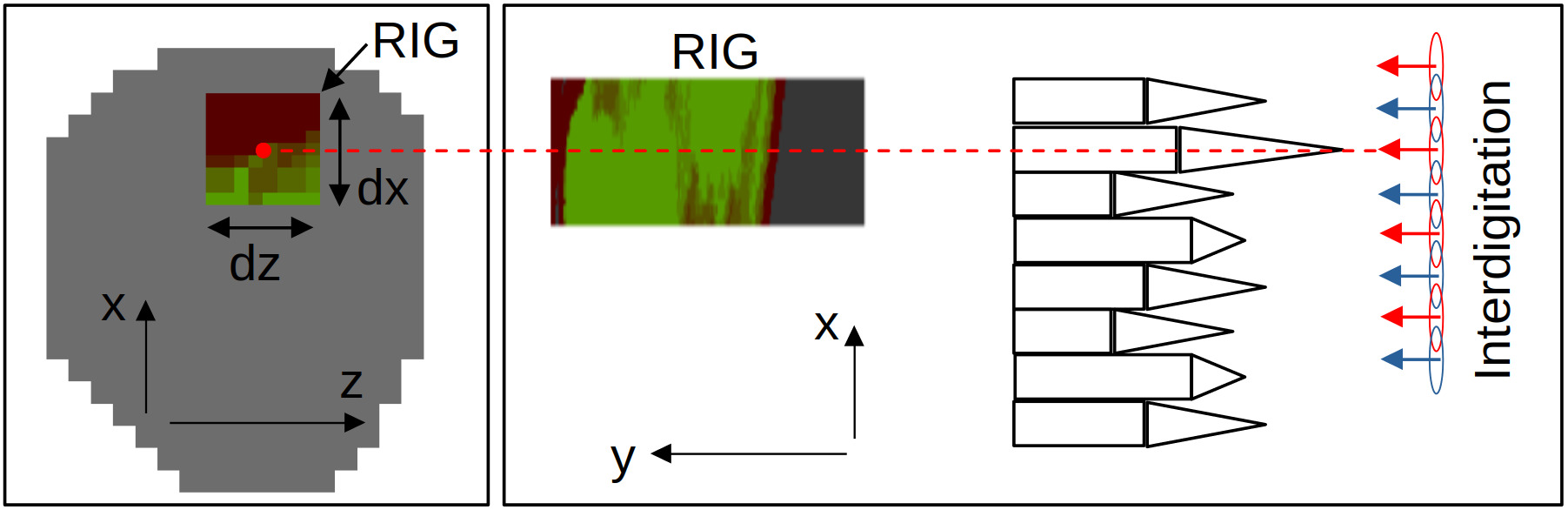}
\caption{Simulation showing restricted influence grid.}
\label{fig:sim_with_restricted_grid}
\end{figure}

\begin{figure}[H]
\centering
\includegraphics[width=4.0in]{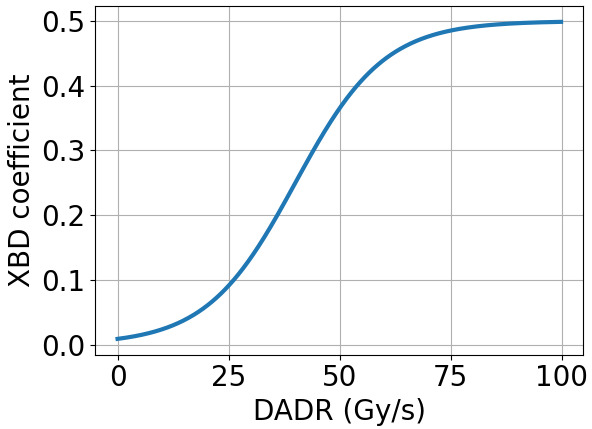}
\caption{Plot of XBD coefficient vs DADR showing a partial advantage at 40 Gy/s, and a full advantage at 100 Gy/s, with the steepest gradient at 40 Gy/s.}
\label{fig:xbd_plot}
\end{figure}

\appendix
\section{IMPT Dose Optimization}
\label{sec:TraditionalIMPT}

Traditional IMPT optimization consists of solving the problem
\begin{equation}
\label{eq:tradimpt1}
\substack{\text{argmin} \\ \dub \in \Real^{N_s}} f \left( \dub \right)
\end{equation}
where $\dub$ are the spot weights, $N_s$ is the number of spots, and
\begin{equation}
\label{eq:tradimpt2}
f \left( \dub \right) = \sum\limits_n p_n f_n^D \left( \dub \right)
\end{equation}
where $f$ is the overall objective function, and $f_n^D$ are the individual dose objectives with relative weights (or ``penalties'') $p_n$.
The solution to the optimization problem is usually bounded by upper and lower limits on the spot weights (e.g. positivity or minimum MU constraints).

Many different dose objectives can be defined.
For example, one common one is the squared deviation objective
\begin{equation}
\label{eq:sqdev}
f_{\text{sq dev}} = \frac{1}{N_v} \sum\limits_{i \in S} \left( d_i - \hat{d} \right)^2
\end{equation}
where $S$ is the set of voxels within a given structure (e.g. tumor, heart, lungs, etc.), $N_v$ is the number of voxels in $S$, $d_i$ is the dose to voxel $i$, and $\hat{d}$ is the prescribed dose.
This objective penalizes the overall objective function every time a voxel's dose deviates from the prescription, with larger deviations leading to larger penalties.
A nice summary of many of the most common dose objectives can be found in \cite{wieser2017}.

The dose to a given voxel, $d_i$, requires the dose influence matrix, $D_{ij}$, which gives the dose per particle to voxel $i$ due to spot $j$, to be known, i.e.
\begin{equation}
\label{eq:di}
d_i = \sum\limits_{j} w_j D_{ij}
\end{equation}
where $w_j$ is the weight of, or number of particles in, spot $j$.
$D_{ij}$ is typically calculated using a MC simulation \cite{souris2016} or an analytical dose engine \cite{xjia2012}, with MC being preferable.
While this calculation can be CPU intensive, it is not in general problematic given modern computing power, and it only needs to be performed once, since $D_{ij}$ is a constant in this context.
This sort of optimization problem usually represents a convex problem and, once $D_{ij}$ is known, it can be solved fairly easily using standard optimization techniques.

\end{document}